\documentclass{article}

  \usepackage[final,nonatbib]{neurips_2024}

\usepackage[utf8]{inputenc} % allow utf-8 input
\usepackage[T1]{fontenc}    % use 8-bit T1 fonts
\usepackage{hyperref}       % hyperlinks
\usepackage{url}            % simple URL typesetting
\usepackage{booktabs}       % professional-quality tables
\usepackage{amsfonts}       % blackboard math symbols
\usepackage{nicefrac}       % compact symbols for 1/2, etc.
\usepackage{microtype}      % microtypography
\usepackage{xcolor}         % colors

\usepackage[sorting=none]{biblatex}
\bibliography{hyperrefs}

\usepackage{tcolorbox}
\tcbuselibrary{listingsutf8} % This enables better control over the box

\title{Position Paper: Model Access should be a Key Concern in AI Governance}

\author{Edward Kembery \\
  ERA Fellow \\
  \texttt{edwardkembery@outlook.com} 
  \And
  Ben Bucknall \\
  Centre for the Governance of AI \\
  \And
  Morgan Simpson \\
  ERA }

\begin{document}

\maketitle

\begin{abstract}
   The downstream use cases, benefits, and risks of AI systems depend significantly on the access afforded to the system, and to whom. However, the downstream implications of different access styles are not well understood, making it difficult for decision-makers to govern model access responsibly. Consequently, we spotlight Model Access Governance, an emerging field focused on helping organisations and governments make responsible, evidence-based access decisions. We outline the motivation for developing this field by highlighting the risks of misgoverning model access, the limitations of existing research on the topic, and the opportunity for impact. We then make four sets of recommendations, aimed at helping AI evaluation organisations, frontier AI companies, governments and international bodies build consensus around empirically-driven access governance.
\end{abstract}

\section*{Introduction}

The rapid development of general-purpose AI has led to discussions and significant legislative efforts focused on governing the technology to minimise societal harm while promoting objectives such as economic growth and respect for diverse cultural norms \cite{AISummitNovember2023, USAIEO, EUAIAct, anderljung2023frontierairegulationmanaging}. A key question has concerned which aspects of models to make available and to whom \cite{shevlane2022structured, eiras2024near, seger2023opensourcinghighlycapablefoundation, kapoor2024societalimpactopenfoundation}. Access provisions can vary: some model developers might offer the public a chat interface, while others may share model weights or training data. Access can also be distributed differently between different stakeholders, including internal company members, regulators, AI safety organisations, third-party researchers, and the public. These decisions form the basis of governing model access. 

Poor model access governance could have serious consequences. Insufficiently cautious information release could increase risks \cite{seger2023opensourcinghighlycapablefoundation, davidevan, kapoor2024societalimpactopenfoundation}, while overly restrictive policies could centralise power or stifle critical research \cite{fbOpenSource, mozillaJointStatement, longpre2024safeharboraievaluation}. Experts have argued that both risks may become more likely over time, and could be difficult to reverse \cite{seger2023opensourcinghighlycapablefoundation, mozillaJointStatement}. Failing to establish good governance principles early could lead to new risks, such as those from unsafeguarded dangerously powerful publicly available models, which would be more challenging to mitigate \cite{davidevan}.

Despite this urgency, decision-makers in frontier AI companies and governments are ill-equipped to govern model access due to limited research, lack of empirical data, and inadequate conceptual clarity. Significant legislation acknowledges the need for risk-proportional model release \cite{rsp}, but no clear guidance exists on how deployers should manage different access styles \cite{USAIEO, EUAIAct}. There also remains substantial uncertainty regarding the use cases afforded by different access styles, and the real-world impacts they could have on various stakeholders \cite{carnegieendowmentBeyondOpen}. Resolving these uncertainties will require the work of a dedicated field aimed at gathering empirical data on these outstanding questions.

We consequently spotlight "Model Access Governance", an emerging field focused on answering the questions needed for organisations and governments to make responsible, evidence-based access decisions. As a domain of AI Governance research \cite{dafoe2018ai}, this field would develop policies, norms, laws, and institutions to determine who accesses which features of AI models and under what conditions. We argue that the risks of mismanagement are significant, current understanding of different model access styles and their implications is limited, and improving this understanding through further research could improve decision-making. We consequently make four recommendations to improve understanding around key issues in Model Access Governance, namely that: i) AI organisations extend evaluations across different access styles, ii) frontier AI companies commit to following responsible best practices, (iii) governments work to coordinate research on downstream implications, and iv) international organisations work to build consensus around empirically-driven access governance. Two appendix sections then set out six priority areas for future research and review existing work related to Model Access Governance.

\section{Defining Model Access Governance}

\textbf{Model Access}: A term describing the manner in which individuals, organisations, and systems can interact with or inspect different aspects of an AI model. This involves three components (see Fig. 1).

\textbf{Model Aspects}: AI \textit{models} are comprised of components including code and weights \cite{basdevant2024frameworkopennessfoundationmodels}. They serve as the foundation of AI \textit{systems}, which include infrastructure such as GPUs and user interfaces \cite{basdevant2024frameworkopennessfoundationmodels, EUAIAct}. Developers must decide which model components and broader information---such as training data or outputs---to share with different groups. We refer to these components as \textit{model aspects}.

\textbf{Access Styles:} When a developer provides access to a model aspect with specific permissions or usage restrictions, we term this an access style. Bucknall and Trager (2023) outline key access styles, including \textit{sampling, inspecting, fine-tuning,} and \textit{modifying} \cite{benbucknallsafety}. However, not all model aspects support all access styles, and different or future models may enable new ones. Thus, existing taxonomies should be viewed as provisional.

\textbf{Access Groups:} An AI developer might decide to make particular aspects of a model or particular access styles available to particular groups with distinct needs, permissions, goals, or capabilities, referred to here as \textit{access groups}. Access groups might include intracompany groups, governments and AISIs, trusted third-party auditors and industry partnerships, research communities, and the general public. 

\begin{figure}[h]
    \centering
    \includegraphics[width=1\linewidth]{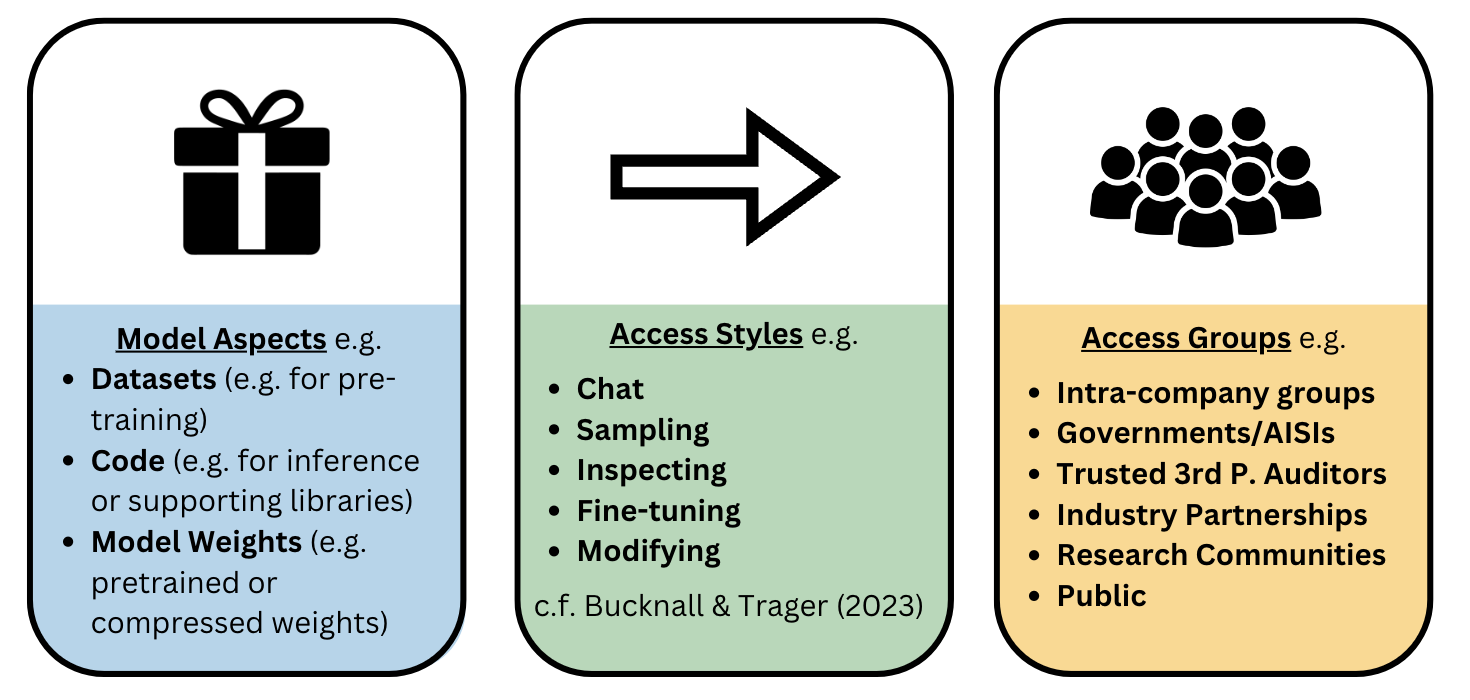}
    \textit{\caption{The Three Elements of Model Access}}
    \label{fig:enter-label}
\end{figure}
\vspace{-2mm} % Reduces vertical space between image and text

\textbf{Model Access Governance}: Governing model access refers to the process of making decisions as to which \textit{access groups} should have what \textit{styles of access} to which \textit{model aspects} under what circumstances. Model Access Governance refers to an emerging research field looking to systematise and build on existing research to build evidence and develop concepts, policies, norms, laws, and institutions that help actors make these decisions better achieve diverse societal objectives. 

For further information about open problems in Model Access Governance and a broader review of related work, see the Appendix. 

\section{The Case for Developing Model Access Governance}

Our case for Model Access Governance is based on three propositions. i) Misgoverning model access could have serious consequences; ii) there is a lack of expert consensus about how to govern model access; and iii) insights and evidence provided by further targeted research can help decision-makers govern model access more effectively. This section expands on each proposition in turn.

\subsection{Misgoverning model access could have serious consequences}

Whilst governing model access incautiously could lead to societal harms, to do so too cautiously could create significant opportunity costs. In the first case, it is already well established that powerful AI systems might have the potential to cause significant societal harms, whether via misuse or accident \cite{anderljung2023frontierairegulationmanaging, weidinger2021ethical}. However, miscalibrated access governance might make these risks significantly worse, for three main reasons:

\begin{itemize}
    \item \textbf{Incautious access policies could make models easier to jailbreak and augment}. Most leading contemporary models are straightforward to jailbreak using text-based commands \cite{aisiAdvancedEvaluations}. However, future models might be significantly more robust to attacks afforded by some styles of access than others \cite{openAI, grayswanAboutGray}. Releasing more information about these more powerful models--for instance, internal chains of thought \cite{openAI}--might make them more vulnerable to jailbreaks and enhancements, increasing the risk of more frequent and serious harm \cite{qi2023finetuningalignedlanguagemodels}. Furthermore, if future models are not robust, it might be possible to use unsecured legacy models could be used to jailbreak them \cite{fang2024teamsllmagentsexploit}. In effect, this would allow malicious actors to `bootstrap' their capabilities up to the frontier. 
    \item \textbf{Miscalibrated access could lead to unsecured models spreading globally, with no way to retract them}. Once models are made public and downloadable, they can be freely shared worldwide, expanding the geographic risk. A poor access decision made \textit{anywhere} creates risks \textit{everywhere}. It also extends the time frame of the risk: a single miscalibrated decision could pose threats \textit{indefinitely} \cite{seger2023opensourcinghighlycapablefoundation, kapoor2024societalimpactopenfoundation}.
    \item \textbf{Miscalibrated access could reduce decision-makers' insight into evolving risks}. When models are made publicly available and downloadable, they can be shared widely and without oversight, diminishing decision-makers' understanding of how the models are used and by whom \cite{seger2023opensourcinghighlycapablefoundation, kapoor2024societalimpactopenfoundation}. This lack of insight may impair their ability to gauge the capabilities of malicious actors or the likelihood of accidents, making them less prepared to address these issues. Paradoxically, the initial ``openness'' might reduce overall transparency about model use, thereby increasing risks.
\end{itemize}

On the other hand, AI systems present the potential to deliver substantial benefits. Miscalibrated model access governance could fail to realise these, incurring opportunity costs in three ways:

\begin{itemize}
    \item \textbf{Miscalibrated access might lead to underutilisation of valuable use cases}. Experts are still uncertain about the marginal utility of different access styles for various use cases and their economic value \cite{acemoglu2024simple, fbOpenSource}. However, restricting access styles for certain groups could significantly limit their ability to develop novel AI applications without substantially increasing risk. Such unnecessary limitations could result in a significant opportunity cost.
    \item \textbf{Miscalibrated access norms could lead to a more unequal distribution of decision-making power over advanced AI}. Sharing specific information about AI systems with certain groups might enable them to gain more benefits than others. For instance, restricting styles of access that allow groups to fine-tune models to improve their performance on low-resource languages \cite{pipatanakul2023typhoonthailargelanguage} might disproportionately affect users who rely on a model's low-resource language capabilities \cite{blasi2021systematic}. Although the impact of access styles on AI value creation and distribution is uncertain, overly centralising decision-making and benefits could represent a significant opportunity cost and potentially lead to risks like the abuse of regulatory power.
     \item \textbf{Miscalibrated access could delay the development of safety research}. Research shows that safety researchers need significant access to AI systems to conduct effective studies \cite{benbucknallsafety, casper2024blackboxaccessinsufficientrigorous}. If current models are low-risk but future ones may pose greater risks, the present could be a critical period for safety research. Missing this opportunity might lead to substantial opportunity costs, resulting in increased risks from more advanced AI systems in the future.
\end{itemize}
   
\subsection{Experts are currently uncertain about how to govern model access and unlikely to get it right by default}

Experts are currently uncertain about how to govern model access \cite{carnegieendowmentBeyondOpen, demosOpenSourcing}. In particular, we note that the existing work is crucially limited by three features: lack of data, lack of adequate concepts, and limited scope. 

\begin{itemize}
    \item \textbf{Experts lack sufficient data to support empirical decision-making}. Governing model access involves understanding the net marginal benefits for different parties under certain technological assumptions and the potential downstream impacts.\footnote{``Technological assumptions'' here might include questions such as whether the model is frontier-level, or whether more powerful models are already accessible in the suggested way.} Despite a few notable exceptions \cite{casper2024blackboxaccessinsufficientrigorous, benbucknallsafety}, empirical evidence as to potential risks and benefits is generally lacking. As a result, foundational work in model access governance often relies on theoretical risks and benefits rather than concrete data. To move beyond speculation and make empirically driven decisions, experts need better, more specific evidence on the risks and benefits of various access styles \cite{carnegieendowmentBeyondOpen}. 

    \item \textbf{Experts lack adequate concepts}. AI models are complex, and existing concepts often equip decision-makers with the wrong intuitions. For instance, models are referred to as `open-source' \cite{demosOpenSourcing, fbOpenSource}, even though the Open Source Initiative has yet to propose an official definition of the term \cite{OSIOpensource}, which may have little to do with open source software \cite{carnegieendowmentBeyondOpen}. Confusingly, the term has been used to refer to many different types of access regime \cite{eiras2024near}. Papers aiming to replace `open vs closed' with a `spectrum of access' \cite{solaiman2023gradientgenerativeairelease, carnegieendowmentBeyondOpen} have been helpful, but risk implying that access to different components of different AI models supports downstream use cases in a way that scales linearly or generalises across domains. Nor are access styles necessarily distinct, since users may be able to extract information relating to one component of a model by accessing another \cite{carlini2024stealing}. Inadequate concepts serve no one: decision-makers will require appropriate concepts to navigate decisions responsibly \cite{carnegieendowmentBeyondOpen, basdevant2024frameworkopennessfoundationmodels}. 

    \item \textbf{Existing research is limited in scope}. Significant work has begun to scope out fundamental issues in Model Access Governance such as potential risks and benefits of different access styles \cite{seger2023opensourcinghighlycapablefoundation, eiras2024near, kapoor2024societalimpactopenfoundation} (see Appendix 2 for an extended literature review). However, this research has tended to focus on public access, neglecting discussions about other stakeholders like internal actors, third party groups, AISIs, and governments. It has emphasized the importance of features like model weights, overlooking the potential for new combinations of access styles. Other important questions, like how trade-offs between risks and benefits should be navigated under uncertainty, have also been overlooked.
\end{itemize}

We think that model access governance is unlikely to go well by default for three reasons. First, companies developing may lack the incentives to govern model access responsibly, especially if responsible governance contradicts fiduciary responsibilities \cite{economistFirmsMustnt}. Second, AI regulation has proved controversial and difficult to pass \cite{scottkohler}, and lack of clarity could lead to further stagnation, or the passing of unhelpful legislation that is then difficult to retract. Finally, there is no international consensus on how to govern model access, meaning that different countries and governments are likely to take different approaches \cite{eiras2024near}, undermining one another's jurisdictional integrity. 

\subsection{Further research will help decision-makers govern model access more effectively}

Further research does not always make problems easier to solve. We believe that building Model Access Governance would support better model access governance in practice for three reasons:

\begin{itemize}
    \item \textbf{Uncertainty is high enough for evaluations to have an impact}. When experts are uncertain, they might default to ideological positions or legacy concepts. Aspiring towards empirical evidence (for example, as to the marginal uplift from different access styles for particular tasks) might not only support accurate scenario analysis and decision-making, it would help researchers recognise shared goals. 
    \item \textbf{The necessary institutions are already in place}. Most crucial empirical inputs for governing model access can be done by industry organisations, AISIs, governments and international bodies, many of whom are already taking modest action in this direction (see Appendix 2). These organisations are well-positioned to contribute to model access governance: for instance, by extending existing risks assessment processes across various access styles (see Recommendation 1).  
    \item \textbf{AI research has a good track record of affecting decision-making}. AI research has repeatedly seen academic ideas like compute governance inform national legislation \cite{sastry2024computing, USAIEO}. Structured access for AI systems already a good example of an academic idea \cite{shevlane2022structured} that has shaped research directions and policy discourse \cite{solaiman2023gradientgenerativeairelease, seger2023opensourcinghighlycapablefoundation}. A strong research field would be well-placed to influence practical decision-making.
\end{itemize}

\section{Recommendations}
\vspace{-2mm}
While governing model access effectively is challenging, it is also feasible. Research on these critical open problems could significantly improve the ability of decision-makers to make informed, responsible decisions. To support this, we make four sets of recommendations, aimed at AI evaluation organisations, frontier AI companies, governments and international bodies respectively. We divide each section into short-term, medium-term and long-term recommendations, to differentiate urgent proposals from those which might require longer timelines to achieve. 

\subsection{AI evaluations should be extended to different styles of access}

Despite their shortcomings \cite{burden2024evaluatingaievaluationperils, mukobi2024reasonsdoubtimpactai}, AI systems evaluations are one of the best ways to determine empirical evidence as to the capabilities of models under different scenarios \cite{aisiAdvancedEvaluations}. Building this capacity in industry and third-party organisations like AI Safety Institutes will provide decision-makers with the empirical evidence necessary to make responsible, confident decisions.

\begin{itemize}
    \item \textbf{Short-term (6 months):} Evaluate how different access styles impact model capabilities on harmful benchmarks, robustness to jailbreaking, and the effectiveness of post-training enhancements. Organisations should flag discrepancies and, if needed, report concerns to appropriate decision-makers.

    \item \textbf{Medium-term (12 months):} Assess how model properties change with user interaction. For example, compare expert prompt engineers to non-experts under various conditions. Evaluate models across diverse user cases, from low-skill actors to sophisticated ones, and under different resource levels to estimate potential harms or benefits.

    \item \textbf{Long-term (18+ months):} Conduct studies on long-term trends in post-training enhancements and scaffolding that might affect different styles of access. This can strengthen confidence in irreversible access decisions, such as weights release. Consider developing model interfaces which support a broader range of use cases in an reversible style, such as fine-tuning APIs.
\end{itemize}

\subsection{Frontier AI companies should commit to following best practices around responsible access governance}

It is crucial that empirical evidence translates to responsible decisions. Consequently, AI deployers will play a key role in ensuring model access is governed responsibly. Many industry actors already have substantial deployment policies like Responsible Scaling Policies \cite{rsp}, and might extend these to include `Responsible Access Policies' \cite{kembery2024aisafetyframeworksinclude} which stipulate how access decisions should be managed.

\begin{itemize}
   \item \textbf{Short-term (6 months):} Frontier AI companies should expand their safety evaluations to include different styles of model access, describing any limitations and mitigation plans \cite{kembery2024aisafetyframeworksinclude}. They should make pre-commitments to provide certain access styles only if deemed low-risk by an appropriate decision channel. They should deepen their collaboration with governments and evaluation organizations to better understand the downstream impacts of model access.

    \item \textbf{Medium-term (12 months):} Frontier AI companies should publish comprehensive policies around access, clearly specifying criteria for roll-forward, roll-back, and non-release decisions. Decision-making processes should be transparent to promote best practices and reduce uncertainty for groups relying on different access styles.

    \item \textbf{Long-term (18 months +):} Frontier AI companies should investigate, support, and consider funding research that enhances empirical understanding of model access regimes on an ongoing basis, such as AI incident reporting mechanisms \cite{carnegieendowmentBeyondOpen}.
\end{itemize}

\subsection{Governments should build the capacity to support effective access governance}

Industry actors may not have the information or the incentives to govern model access responsibly. Consequently, national governments will play a crucial role in supporting model access governance. They should support AI model evaluations, coordinate research around AI impacts, and consider regulation.

\begin{itemize}
    \item \textbf{Short-term (6 months):} Governments should support national AI safety organisations and internal evaluation structures to evaluate risks from different styles of model access. They should consult with frontier AI companies to build an informed consensus around regulation. They should collaborate to build an international consensus (Recommendation 4).
    
    \item \textbf{Medium-term (12 months):} Governments should investigate the impact of different access styles on the AI ecosystem. They might coordinate further third party research around downstream effects like economic uplifts, and cyber-security implications. 
    
    \item \textbf{Long-term (18 months +):} Governments should consider regulation to support responsible model access governance: for instance, by including a requirement that AI companies follow responsible access governance best practices in national legislation. They should consider how to best assist those affected by access governance, including victims of access retraction, or disadvantaged by access trade-offs. 
\end{itemize}

\subsection{International bodies should strive for a global consensus on governing model access}

If access is misgoverned \textit{anywhere}, it could create risks \textit{everywhere}. Accordingly, international organisations like the UN and OECD may play a crucial role ensuring that jurisdictional integrity is preserved and access is managed suitably \cite{carnegieendowmentEnvisioningGlobal}. We expect that this may involve longer timeframes. 

\begin{itemize}
    \item \textbf{Short-term (12 months):} International bodies should platform discussions aiming to build consensus around empirically supported governance among member states. Questions around model access governance may be raised at conventions like AI Safety summits and through channels like the International Scientific Report on the Safety of Advanced AI. 
    \item \textbf{Medium-term (24 months):} International bodies should platform discussions encouraging member states to adopt stipulations for requiring companies within national jurisdictions to comply with access governance best practices. These discussions should aim to identify potential empirical and value-based disagreements.
    \item \textbf{Long-term (36 months +):} International processes should platform discussions supporting commitments around responsible model access governance. These discussions may cover how member states can best distribute the task of evaluating evolving models, or implement verifiable agreements that could be adopted by many nations in future \cite{cha2024towards, wasil2024verification}.
\end{itemize}

We recognise that the future of model access governance is uncertain, and emphasize international bodies should be agile and seek to adapt their policy to reflect the changing technological landscape. 

\newpage

\section*{Acknowledgements}

This research was supported by the ERA Fellowship. The authors would like to thank the ERA Fellowship for its financial and intellectual support, as well as Marius Hobbhahn, Guarav Sett, Lisa Soder, Jack Wadham and Tobin South for their comments and insight. 

%\section{Bibliography}

\printbibliography

\newpage

\appendix
\section{Some Open Problems in Model Access Governance}

This section gestures to six categories of open problems Model Access Governance should prioritize: 

\begin{tcolorbox}[colback=purple!10!white, colframe=brown!70!black]
    \textbf{4.1. Establishing Access Elements}. Decision-makers lack a clear and consistent framework for describing different elements of model access. They need options and terms to describe them which are distinct, feasible, consistent, and precise. For example: 
    \begin{itemize}
        \item How can access elements be defined so as to support effective decision-making?
        \item How can labs ensure decisions to withold access are robust and enforceable?
    \end{itemize}
\end{tcolorbox}   

\begin{tcolorbox}[colback=red!10!white, colframe=red!70!black]
        \textbf{4.2. Evaluating Risks}. The risks of providing different access styles to different groups is currently highly unclear. Decision-makers need accurate estimates of potential uplifts in misuse risks, systemic risks, and risks to privacy given different access regimes. E.g.:
        \begin{itemize}
        \item What is the net marginal uplift in risks from variously capable malicious actors given various styles of model access, under various technological assumptions?
        \item How might different model access regimes affect robustness, create security vulnerabilities, or compromise privacy? 
        \end{itemize}
\end{tcolorbox}

\begin{tcolorbox}[colback=orange!10!white, colframe=orange!70!black]
        \textbf{4.3. Evaluating Benefits}. The benefits of providing different access styles to different groups is currently highly unclear. Decision-makers need accurate estimates of potential uplifts in value to different groups, security, and scientific research given different access regimes. E.g.:
        \begin{itemize}
        \item What is the net marginal uplift in value from AI models for different use cases given various styles of model access, and how might this be distributed differently between different groups?
        \item How might different model access regimes improve the robustness and security of models, improve privacy, or accelerate research into AI systems?
    \end{itemize}
\end{tcolorbox}

\begin{tcolorbox}[colback=yellow!10!white, colframe=yellow!70!black]
        \textbf{4.4. Navigating Trade-offs}. Governing model access responsibly will require value-driven decisions under significant uncertainty. Decision-makers need guidance and principles for weighing evidence, eliciting inputs from stakeholders, and aligning corporate incentives. E.g.:
        \begin{itemize}
        \item How should decision-makers govern model access given high uncertainty or risks which affect different groups asymmetrically?
        \item How should decisions be structured to resist perverse incentives such as (for instance) fiduciary responsibility or pressure from a `race' scenario?
    \end{itemize}
\end{tcolorbox}

\begin{tcolorbox}[colback=green!10!white, colframe=green!70!black]
        \textbf{4.5. Paths to Impact}. Governing model access effectively will require inter-organisational collaboration. Labs, AISIs, governments, third parties and international bodies need clear roles and processes to support best practices in model access governance. E.g.:
        \begin{itemize}
        \item How might frontier AI companies adopt processes for empirically evaluating access styles, and providing them when benefits sufficiently outweigh the risks?
        \item Which organisations are best placed to play which roles in model access governance, and where should capacity be strengthened to support them? 
    \end{itemize}
\end{tcolorbox}

\begin{tcolorbox}[colback=blue!10!white, colframe=blue!70!black]
        \textbf{4.6 Future-Proofing}. Decisions about model access may be difficult to reverse, even as the technological landscape changes substantially. Decision-makers need a clear view of trends in AI and their potential implications for model access governance. E.g.:
        \begin{itemize}
        \item How will model access governance change as models become (for instance) cheaper, more powerful, more robust, smaller, or more valuable for certain use cases? 
    \end{itemize}
\end{tcolorbox}

\section{Related Work}

Significant contributions to Model Access Governance have already been made across diverse fields. By highlighting and presenting this work as a unified field, we aim to emphasize the value of integrating and coordinating this research, and inspire further progress in this area.

\textbf{Structured access, transparency, and information security}. Structured transparency{\textemdash}a framework designed to facilitate collaboration and information sharing whilst addressing privacy concerns \cite{trask2024privacytradeoffsstructuredtransparency}{\textemdash}has been applied to model access decisions around AI systems for several years \cite{shevlane2022structured}. Recent work has inquired as to the information required to do valuable safety research \cite{benbucknallsafety} and how audits might best be structured \cite{longpre2024safe, mokander2023auditing, mokander2022conformity, mokander2021ethics, mokander2023operationalising}. 

At the same time, recent technical work has explored the extent to which these attempts to control information flows are robust: for instance, how easy it is to ``steal'' the weights of a production language model through interactions \cite{tramèr2016stealingmachinelearningmodels}. Early attempts to steal models with high fidelity attacks focused on specific deep neural networks with ReLU activations \cite{251526, milli2018modelreconstructionmodelexplanations}, but more recent work has demonstrated that attacks can succeed against larger models \cite{Wei2020LeakyDS, pmlr-v139-zanella-beguelin21a} and even parts of production models \cite{carlini2024stealingproductionlanguagemodel}. Furthermore, recent theoretical work on information security has suggested that the weights of models developed by frontier AI companies might not be resilient to extraction attacks by sophisticated and state-level actors, suggesting the need for further security in order to govern model access responsibly \cite{nevo2024securing}.

\textbf{Open vs closed-source debate}. Early work related to model access governance transferred a framing of `openness' from earlier software research (e.g. Millar et al., 2022 \cite{millar2022vulnerabilitydetectionopensource}). These concepts and practices have not always transferred seamlessly to the governance of AI models \cite{basdevant2024frameworkopennessfoundationmodels}, and there is as yet no official definition of `open source AI' \cite{OSIOpensource}. Some research has argued that we should move `beyond the open vs closed debate' \cite{carnegieendowmentBeyondOpen} towards `hybrid' access regimes that involve providing access to some aspects of the model but not others \cite{solaiman2023gradientgenerativeairelease}. Nonetheless, several papers have usefully contributed to discussions around model access governance by exploring possible marginal risks from models with widely-available weights \cite{stanford2023open} and possible alternatives to open-weight deployment \cite{kapoor2024societalimpactopenfoundation,seger2023opensourcinghighlycapablefoundation}. Others have made a vigorous case for various definitions of open-source \cite{eiras2024near, NTIARequestforcomment}.

\textbf{Model Deployment Governance}. Frontier AI companies that have produced safety frameworks around AI systems include Anthropic \cite{rsp}, Google DeepMind \cite{deepmindIntroducingFrontier} and OpenAI \cite{OpenAIPreparedness}. However, the extent to which these frameworks are comprehensive or feasible is unclear \cite{alaga2024gradingrubricaisafety}, and they currently do not provide explicit commitments for how model access should be governed such as criteria and standards for the styles of access that should be provided to different groups. Companies that have made explicit comments in favour of `open' access styles include Meta \cite{fbOpenSource}, Mozilla and LAION \cite{NTIARequestforcomment}, but these commitments may not be binding as the technology develops. Encouragingly, a recent update to Anthropic's Responsible Scaling Policy suggested that `developing a tiered access system' was a priority \cite{anthropicRSPupdate}, suggesting that developments to incorporate responsible access governance into existing safety policies may be under consideration. However, as the earlier report noted, ``we are a company that primarily develops proprietary systems, and we don’t have the legitimacy to make claims here about what should or shouldn’t be acceptable in openly disseminated systems'', suggesting the need for more substantial collaboration between frontier AI companies, governments, and third-party auditing and evaluation organisations \cite{anthropicThirdpartyTesting}.

\textbf{Government Practice, Reports and Legislation}. No major legislation currently distinguishes between styles of access explicitly or provides guidance for frontier AI companies making decisions about which aspects or style to provide to particular groups. However, preliminary steps have already been taken towards governing model access responsibly. In Europe, for instance, Article 53 of the EU AI Act indicates some stipulations as to the variety of information providers of general purpose AI models must make available to different parties \cite{EUAIAct}. However, there is some ambiguity as to the nature of these requirements (e.g. what a ``sufficiently detailed summary about the content used for training'' might look like for contemporary large language models). There is also significant ambiguity as to what information about models is to be provided to different third-parties for performing model evaluations (Art. 92). Furthermore, the definition of ``general-purpose AI models with systemic risks'' in Article 51 does not indicate how the nature of access provided to the model might contribute to the risk profile of the model \cite{EUAIAct}. Meta recently chose not to open-source a model in the EU, citing ``regulatory uncertainty'' \cite{nationaltechnologyMetaWithholds}. 

The US Government has also taken steps towards governing model access. Section 4.6 of the Executive Order tasked the Secretary of Commerce with soliciting feedback ``from the private sector, academia, civil society, and other stakeholders through a public consultation process on the potential risks, benefits, other implications, and appropriate policy and regulatory approaches related to dual-use foundation models for which the model weights are widely available'' \cite{USAIEO}. The National Telecommunications and Information Administration (NTIA)  received 332 comments from their Request for Comment published in February 2024 \cite{NTIARequestforcomment}. In July they published an explicit statement in favour of open-source \cite{ntiaNTIASupports}. However, their report on the topic, `Dual-Use Foundation Models with Widely Available Model Weights', recognises both risks and benefits, and recommends that the US government builds capacity and considers alternatives as the technology develops \cite{ntiaPaper}. Beyond the US and EU, legislation is currently meagre and reference to model access governance is limited.

\end{document}